\documentclass[intlimits,twoside,a4paper]{article}

\usepackage{amsmath,amssymb}
\usepackage{graphicx}
\usepackage{subfig}
\usepackage[T2A]{fontenc}
\usepackage[cp1251]{inputenc}

\usepackage[eqsecnum]{cmpj3}

\usepackage{bm}

\issue{2020}{23}{1}{13701}
\doinumber{10.5488/CMP.23.13701}
\title[Phase stability, electronic and magnetic properties]%
{Theoretical study of phase stability, electronic and magnetic properties of Rh$_2$CrGe$_{1-x}$Al$_x$ ($x = 0$, $0.25$, $0.50$, $0.75$ and $1$) Heusler alloys by FP-LAPW method}

\author[M. Kadjaoud \textsl{et al.}]{M. Kadjaoud\refaddr{label1}, M. Mokhtari\refaddr{label2,label3},
        L. Djoudi\refaddr{label1,label3}, M. Merabet\refaddr{label1,label3}, S. Benalia\refaddr{label1,label3}, D. Rached\refaddr{label1}, R. Belacel\refaddr{label1}, F. Zami\refaddr{label4}, F. Dahmane\refaddr{label3}}
\addresses{
\addr{label1} Magnetic Materials Laboratory, Physics Department, Sidi-Bel-Abbes University, Algeria
\addr{label2} Universit\'e des Sciences et de la Technologie d'Oran Mohamed Boudiaf, USTO-MB, LEPM, BP 1505, \\ El M' Naouar, 31000 Oran, Algeria
\addr{label3} D\'epartement de SM, Institut des Sciences et des Technologies, Centre Universitaire de Tissemsilt, \\ 38000 Tissemsilt, Algeria 
\addr{label4} Condensed Matter and Sustainable Development Laboratory, Physics Department, Sidi-Bel-Abbes University, Algeria
}

\date{Received July 12, 2019, in final form August 29, 2019}

\begin{document}

\maketitle

\begin{abstract}
 First-principle calculations were performed within the framework of the density functional theory (DFT) using FP-LAPW method as implemented in WIEN2k code to determine the structural stability, electronic and magnetic properties of Rh$_2$CrGe$_{1-x}$Al$_x $($x = 0$, $0.25$, $0.50$, $0.75$ and $1$). The   results   showed   that   for   Rh$_2$CrAl and Rh$_2$CrGe, the Cu$_2$MnAl-type  structure  is  energetically more  stable  than  Hg$_2$CuTi-type  structure  at  the  equilibrium volume. The calculated densities of states for Rh$_2$CrAl and Rh$_2$CrGe show half-metallic and nearly half-metallic behavior, respectively. Rh$_2$CrGe$_{1-x}$Al$_x$ ($x = 0.25$, $0.50$, $0.75$) these alloys show a half-metallic character, and these compounds are predicted to be good candidates for spintronic applications.

\keywords Heusler alloys, structural properties, electrical properties, magnetic properties 
%
\end{abstract}

\section{Introduction}
Heusler alloys have been a subject of unprecedented research since their discovery in 1903 by the German engineer F. Heusler \cite{1, 2}. The experimental works showed that the majority of Heusler alloys are ferromagnetic ordered in stoichiometric compositions~\cite{3,4}. In the very recent past, Heusler compounds have received a great deal of interest, due to their potential applications in many fields. In particular, magnetic Heusler alloys are mostly used in spin-based electronic devices \cite{5}, thermoelectric \cite{6,7} and superconductors \cite{8}. Both high-spin polarizations and half-metallicity are lately considered as the key factors in such type of materials \cite{9, 10}. It is well known that most of Heusler alloys have the spin polarization near Fermi level which is as high as $100\%$ \cite{11}. Some of the Heusler alloys have only one spin channel for conduction at the Fermi level, while for the other spin channel, a semiconducting gap appears between  the valence band and the conduction band~\cite{9,12,12a}.

The possibility of tuning the electronic and magnetic properties, especially the spin polarization, many of $X_2YZ$ full, half and inverse Heusler compounds have been explored. 
For example, Rh$_2YZ$ represent an interesting family in the world of Heusler compounds which has been intensively studied during the recent decades. The crystal structure and magnetic properties have been determined for a new series of compounds of the form Rh$_2T$Sn for $T$ = Mn, Ni, or Cu \cite{11}. In their work, M. Pugaczowa-Michalska and A. Jezierski \textsl{et al.} studied the magnetic properties of Rh$_2T$Sn Heusler alloys ($T$=Mn, Fe, Co, Ni and Cu)~\cite{13}. Mohammed El Amine Monir \textsl{et al.} studied half-metallic ferromagnetism in the novel Rh$_2$-based full-Heusler alloys Rh$_2$Fe$Z$ ($Z$ = Ga and In)~\cite{14}. Rh$_2$CrSb, Rh$_2$MnBi, Rh$_2$MnAl, Rh$_2$CrAl and Rh$_2$CrIn, another series of full heusler alloys, were studied by first-principles computational methods~\cite{15}.

In the present work, we focused our study on the structural, electronic and magnetic properties of Rh$_2$CrGe$_{1-x}$Al$_x$ ($x = 0$, $0.25$, $0.50$, $0.75$) compounds. The  rest  of  the  paper  is  arranged as  follows: section~\ref{sec2}  includes  computational details  and the  method  of  calculation,  section~\ref{sec3}  is  devoted  to  the results and discussion, and section~\ref{sec4} is a summary of our conclusions.

\section{Computational method}\label{sec2}

We have carried out first principles calculations using the full potential linear augmented plane wave (FP-LAPW) methods \cite{16} as implemented in the WIEN2k code \cite{17}  in the framework  of the density functional theory (DFT) \cite{18, 19} within the generalized gradient approximation (GGA-PBE) \cite{20}. In this method, the space is divided into non-overlapping muffin-tin (MT) spheres separated by an interstitial region. In this context, the basic functions are expanded in combinations of spherical harmonic functions inside the muffin-tin spheres and Fourier series in the interstitial region. In the calculations, the Rh ($4d^85s^1$), Al ($3s^23p^1$), Cr ($3d^54s^1$) and Ge ($3d^104s^24p^2$) states are treated as valence electrons.

The valence wave functions inside the MT spheres are expanded in terms of spherical harmonics up to $l_\text{max} = 10$. We set the parameter $RMT$. $K_\text{max} = 7$ (where $RMT$ is the average radius of the MT spheres and $K_\text{max}$ is the largest reciprocal lattice vector used in the plane wave expansion). The magnitude of the largest vector in charge density Fourier expansion ($G_\text{max}$) was $14 $~(a.u.)$^{-1}$. Both the plane wave cut-off and the number of   k-points were varied to ensure total energy convergence. Our calculations for valence electrons were performed in a scalar-relativistic approximation, while the core electrons were treated relativistic. The self-consistent calculations were considered to converge only when the calculated total energy of the crystal converges to less than $10^{-4}$~Ry. 

In order to simulate the Rh$_2$CrGe$_{1-x}$Al$_x$ ($x = 0.25$, $0.50$, $0.75$) quaternary alloy, we generate a supercell with 16 atoms. For $x = 0.25$, we substituted one atom of Ge by one  atom of Al; for $x = 0.50$, we substituted two atoms of Ge by two  atoms of  Al; for $x = 0.75$, we substituted three atoms of Ge by three atoms of  Al.

\section{Results and discussion}\label{sec3}
\subsection{Structural properties}

\begin{figure}[!t]
	\centering
	\subfloat[]{{\includegraphics[width=10cm]{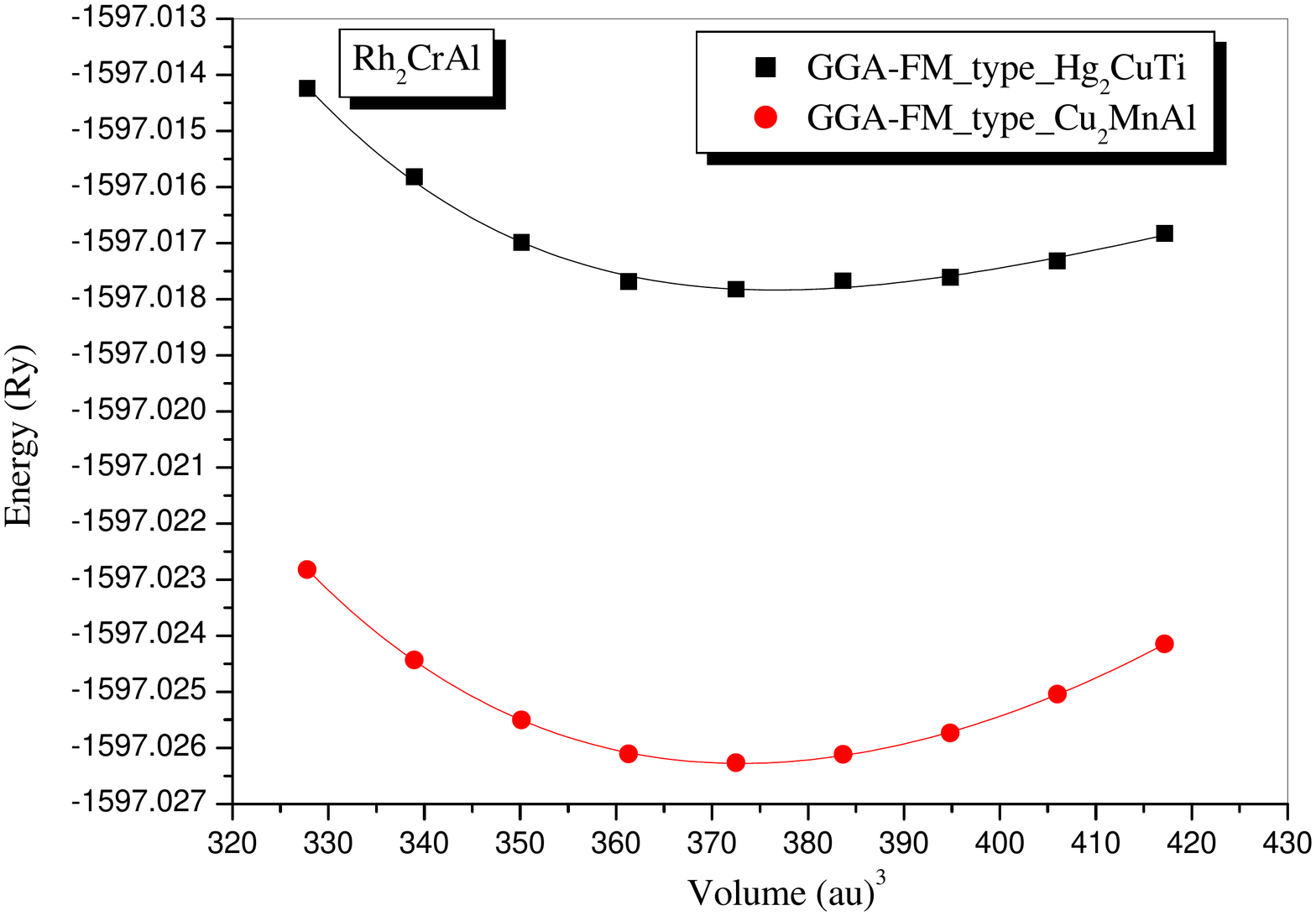} }}%
	\qquad
	\vspace{-4mm}
	\subfloat[]{{\includegraphics[width=10cm]{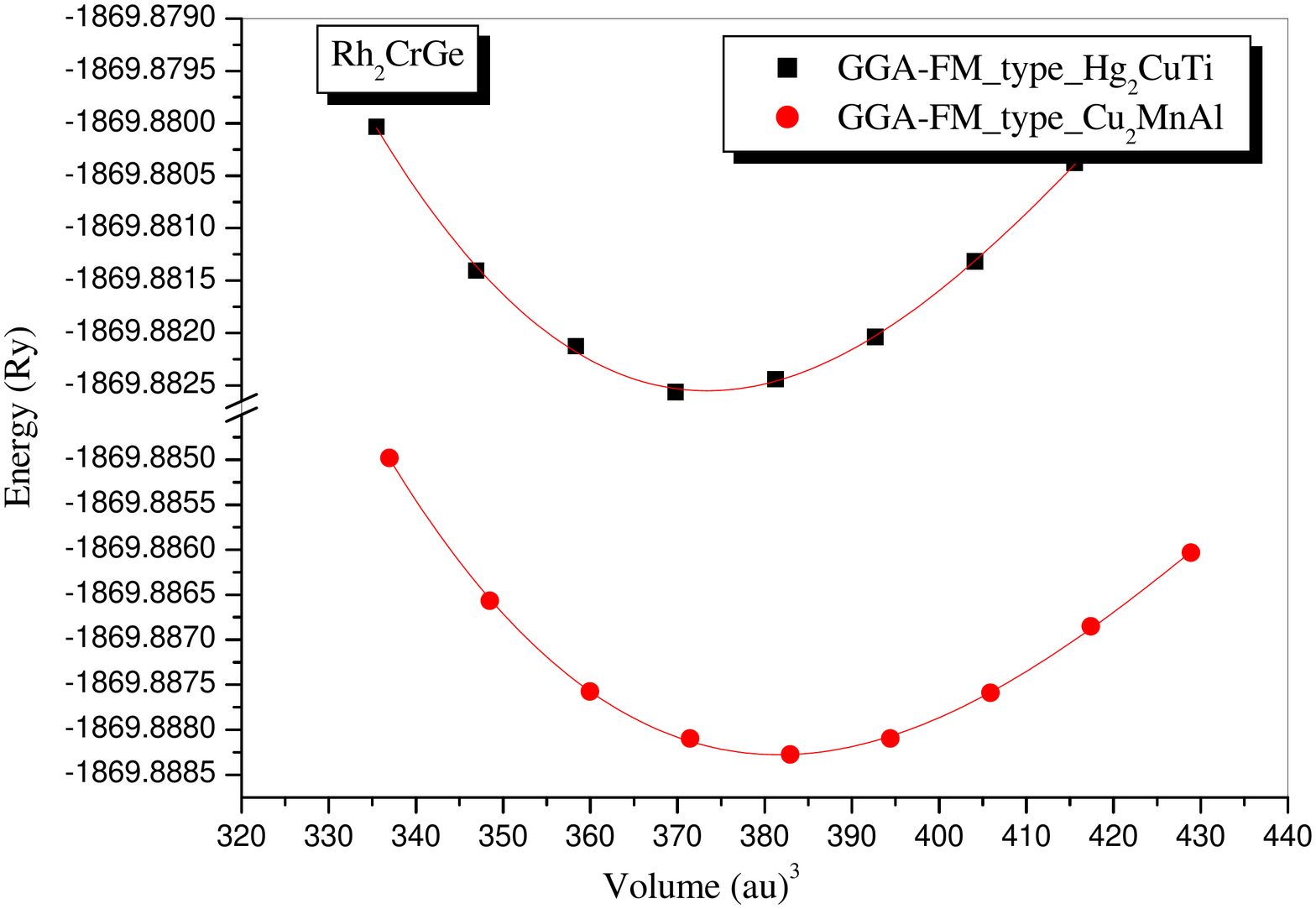} }}%
	\caption{(Colour online) Variation of the total energy as a function of the unit-cell volume of Rh$_2$CrGe, Rh$_2$CrAl for both Hg$_2$CuTi and Cu$_2$MnAl type structure.}%
	\label{fig1}%
\end{figure}

Heusler alloys can be classified into two main groups, namely, half-Heusler $XYZ$ alloys and full-Heusler $X_2YZ$ alloys. Here, $X$ and $Y$ denote transition metal elements, and $Z$ is a $sp$-element. Full Heusler $X_2YZ$ alloys generally have two types of structures \cite{21, 22}. The first one is $L21$ (regular cubic phase, prototype Cu$_2$MnAl in which the two $X$ atoms occupy A $(0,0,0)$ and C $(1/2, 1/2,1/2 )$ positions, and $Y$, $Z$ atoms occupy B  $(1/4,1/4,1/4)$ and D $(3/4,3/4,3/4)$ positions, respectively. The second one is $XA$ (``inverted cubic phase'', prototype Hg$_2$CuTi), in which the two $X$ atoms occupy A $(0, 0, 0)$ and B $(1/4, 1/4, 1/4)$ positions, and $Y$, $Z$ atoms occupy C $(1/2, 1/2, 1/2)$ and D $(3/4, 3/4, 3/4)$ positions, respectively~\cite{22}. According to Luo \textsl{et al.}~\cite{23}, the site preference of the $X$ and $Y$ atoms is strongly influenced by the number of their $3d$ electrons. Those elements with more $3d$ electrons prefer to occupy the A and C sites and those with fewer ones tend to occupy the B sites. As a first step, we studied the phase stability of Rh$_2$CrGe and Rh$_2$CrAl in the two types of structures and we found that Cu$_2$MnAl ($L21$) is more stable as shown in figure~\ref{fig1}.

The total energy as a function of the cell volume curves was fitted to the Murnaghan equation of  \cite{24}, in order to determine the ground state properties
\begin{equation}
E(V)=E_{0}(V)+\frac{BV}{B'(B'-1)}\left[B\left(1-\frac{V_{0}}{V}\right)+\left(\frac{V_{0}}{V}\right)^{B'}-1\right].
\label{moneq}
\end{equation}

Here, $E_0$, $V_0$, $B$ and $B'$ are the equilibrium energy, volume, bulk  modulus and its first derivative, respectively. The equilibrium structural parameters ($a$, $B$ and $B'$) of Rh$_2$CrGe$_{1-x}$Al$_x$ Heusler alloys in both Hg$_2$CuTi and Cu$_2$MnAl phases are given in table~\ref{tab1}.

Note that for the present Heusler alloys [Rh$_2$CrGe$_{1-x}$Al$_x$ ($x= 0.25$, $0.5$, $0.75$)], there are no experimental or theoretical data accessible to get our calculations started. In order to obtain their approximate lattice parameters, we have applied a linear combination of the lattice constants of the Rh$_2$CrGe$_{1-x}$Al$_x$ alloys and the associated concentration $x$ of Al incorporated atom allowing for the so-called Vegard's law \cite{25,26,27}.

Rh$_2$CrGe$_{ 0.75}$ Al$_{0.25}$:  $a= (6.088 \times 0.75)+(6.044 \times 0.25) = 6.077~$\AA,

Rh$_2$CrGe$_{0.5}$Al$_{0.5}$: $a= (6.088 \times 0.5)+(6.044 \times 0.5) = 6.066$~\AA,

Rh$_2$CrGe$_{0.25}$Al$_{0.75}$: $a= (6.088 \times 0.25)+(6.044 \times 0.75) = 6.055$~\AA.

The obtained lattice parameter values for the Rh$_2$CrGe$_{ 0.75}$Al$_{0.25}$, Rh$_2$CrGe$_{0.5}$Al$_{0.5}$  and \newline Rh$_2$CrGe$_{0.25}$Al$_{0.75}$ Heusler alloys are well accorded with Vegard's law values, which are higher by $0.34 \%$, $0.32\%$ and $0.3\%$, respectively. 

By analyzing the present results, we can see the direct proportion between the lattice parameters [obtained by GGA for the Cu$_2$MnAl ($L21$) structure and calculated by Vegard's law] of the three Heusler alloys and the concentration ($x$) of Al atom. This occurs owing to the Ge ($125$~pm) large atomic radius compared to the Al ($118$~pm) radius, respectively. It is clearly seen that the GGA offers greater values than the Vegard's law ones. At the same time, the corresponding bulk modulus values are $199.6756$~GPa, $200.0788$~GPa and $199.0289$~GPa, respectively. We can conclude that Rh$_2$CrGe$_{0.25}$Al$_{0.75}$ is the most compressible full Heusler alloy.

\begin{table}[!t]
\caption{Structural parameters ($a$, $B$ and $B'$) of Rh$_2$CrGe, Rh$_2$CrAl and Rh$_2$CrGe$_{1-x}$Al$_x$ ($x=0.25$, $0.5$ and $0.75$).}
\label{tab1}
\renewcommand{\arraystretch}{1.3}
\vspace{2ex}
\begin{center}

\renewcommand{\arraystretch}{0}
\begin{tabular}{|c|c||c|c|c|c||}
\hline

      Rh$_2YZ$& &&$a$(\AA)&$B$(GPa)&$B'$\strut\\

\hline
\rule{0pt}{2pt}&&&&&\\
\hline
\raisebox{-1.7ex}[0pt][0pt]{Rh$_2$CrGe}
     &\raisebox{-1.7ex}[0pt][0pt] {Cu$_2$MnAl Type} &\raisebox{-1.7ex}[0pt][0pt]{FM} &\raisebox{-1.7ex}[0pt][0pt]{6.936}&\raisebox{-1.7ex}[0pt][0pt]{193.715}&\raisebox{-1.7ex}[0pt][0pt]{4.95}\strut\\
       & &&&&\strut\\
\cline{2-6}

     &\raisebox{-1.7ex}[0pt][0pt]  {Hg$_2$CuTi Type}&\raisebox{-1.7ex}[0pt][0pt] {FM}&\raisebox{-1.7ex}[0pt][0pt]{6.046}&\raisebox{-1.7ex}[0pt][0pt]{227.177}&\raisebox{-1.7ex}[0pt][0pt]{4.155}\strut\\
&&&&&  \strut\\
\hline
\raisebox{-1.7ex}[0pt][0pt]{Rh$_2$CrAl}
      &\raisebox{-1.7ex}[0pt][0pt] {Cu$_2$MnAl Type}&\raisebox{-1.7ex}[0pt][0pt] {FM}& \raisebox{-1.7ex}[0pt][0pt]{6.306}&\raisebox{-1.7ex}[0pt][0pt]{176.087}&\raisebox{-1.7ex}[0pt][0pt]{4.712}\strut\\
 & &&&&\strut\\
\cline{2-6}
        &\raisebox{-1.7ex}[0pt][0pt] {Hg$_2$CuTi Type}&\raisebox{-1.7ex}[0pt][0pt] {FM}&\raisebox{-1.7ex}[0pt][0pt]{6.0468}&\raisebox{-1.7ex}[0pt][0pt]{200.36}&\raisebox{-1.7ex}[0pt][0pt]{4.4762}\strut\\
&&&&&  \strut\\
\hline

Rh$_2$CrGe$_{0.25}$Al$_{0.75}$ & &FM&6.0580&199.0289&4.7142\strut\\

\hline

 Rh$_2$CrGe$_{0.50}$Al$_{0.50}$& &FM&6.0692&200.0788&4.6488\strut\\

\hline

 Rh$_2$CrGe$_{0.75}$Al$_{0.25}$& &FM&6.0804&199.6756&4.7420\strut\\

\hline

\end{tabular}
\renewcommand{\arraystretch}{1}
\end{center}
\end{table}

\subsection{Electronic and magnetic properties}

\begin{figure}[!t]
    \centering
    \subfloat[]{{\includegraphics[width=12cm]{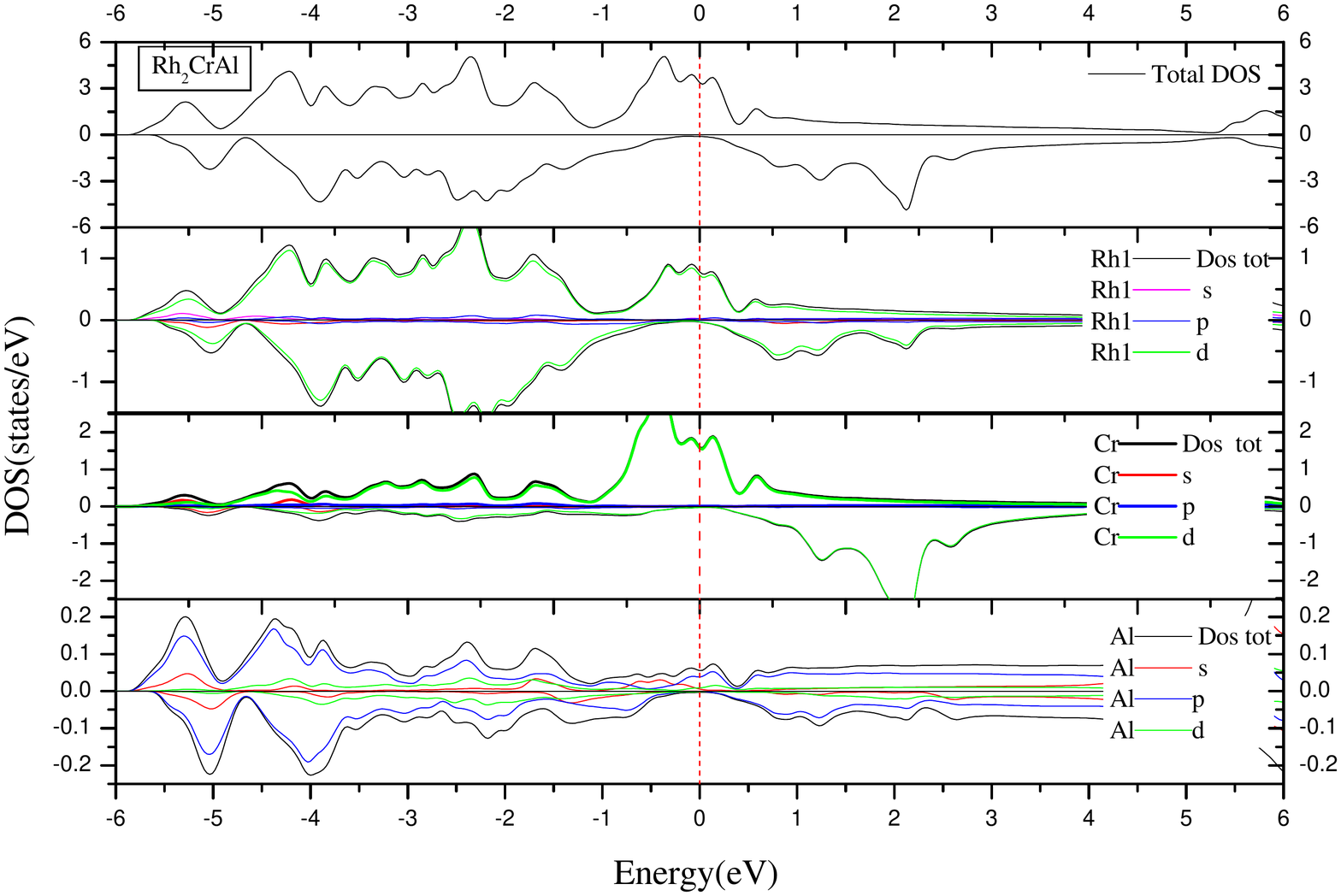} }}%
    \qquad
    \subfloat[]{{\includegraphics[width=12cm]{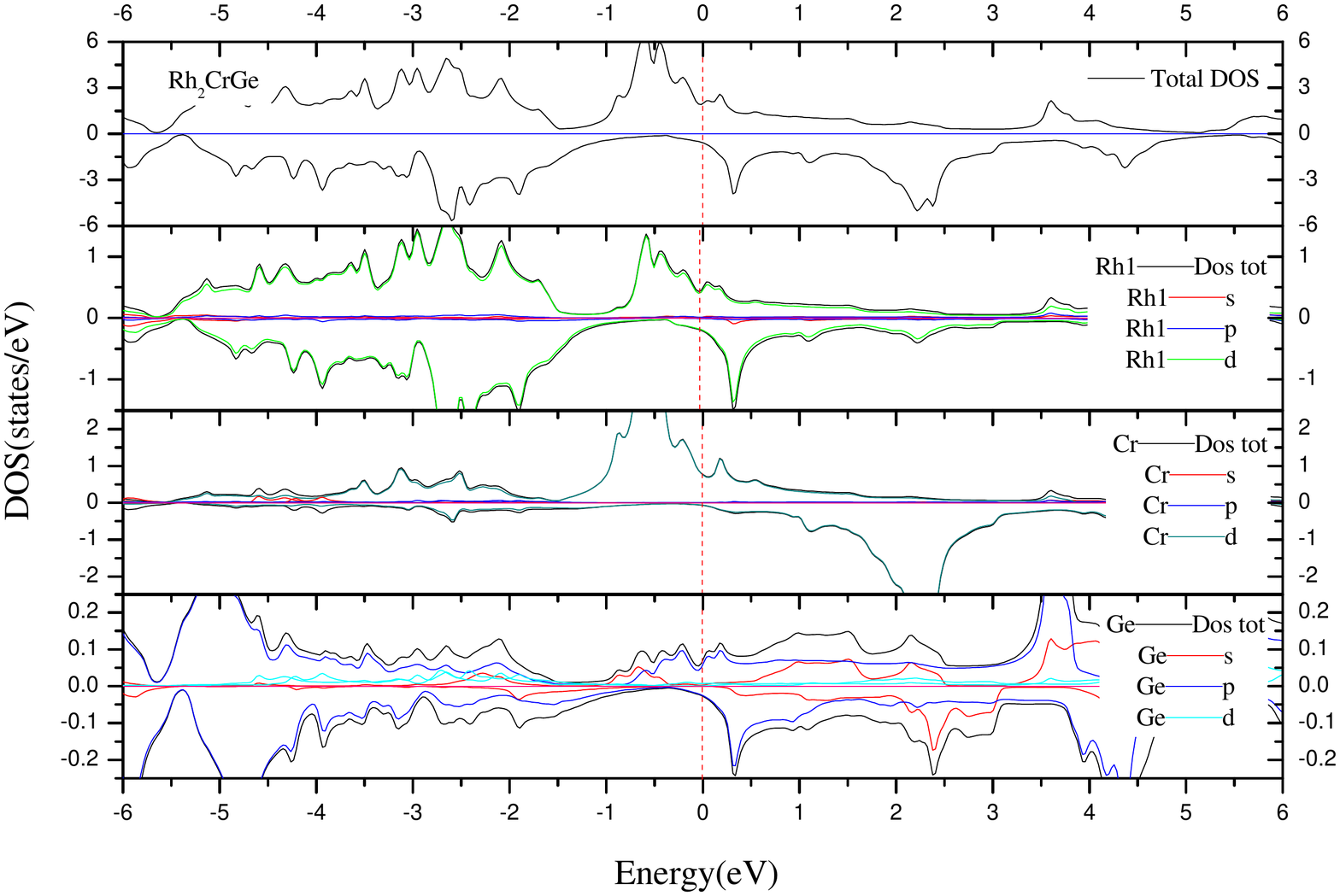} }}%
    \caption{(Colour online) Densities of states  of Rh$_2$CrGe and Rh$_2$CrAl.}%
    \label{fig2}%
\end{figure}

The calculated densities of states for Rh$_2$CrGe and Rh$_2$CrAl are presented in figure~\ref{fig2}, the responsibility of transition metals $3d$-states is very significant in the description of density of sates calculations. The magnetic moments of the materials depend on the interactions of $3d$-band structure with other states \cite{28}. Coey \textsl{et al.} \cite{29} have proposed a large classification scheme for half-metallic ferromagnets: A material with conducting electrons (metallic behavior) at one for the spin channels and an integer moment at $T = 0$ is a good candidate for A type-I half-metallic ferromagnet. This is the phenomenon described by de Groot~\textsl{et al.} \cite{30}. Type-IA half-metals are metallic in spin up, and  semiconducting in spin down; the opposite is true for the type-IB half-metals \cite{31}. For the Rh$_2$CrGe Heusler alloy, the majority-spin bands are metallic due to the overlap between the conduction and valence bands around $E_\text F$, but the minority-spin bands are nearly semiconductors because the conduction band minimum cut little the $E_\text F$ at the $\Gamma$ symmetry point. Thus, Rh$_2$CrGe is nearly half-metallic.
 
For Rh$_2$CrAl full Heusler alloys, we can see that spin up present a metallic behavior while spin down is semiconductor which indicates a half-metallic behavior.
Slater and Pauling \cite{32, 33} had exposed that for a binary magnetic alloy, when we add one valence electron in the compound, this occupies spin-down states only and the total spin magnetic moment decreases by about $1$ $\muup \text B$. For $L21$  full-Heusler, the relation between the total spin magnetic moment in the unit cell $M_\text t$ and total number of valence electrons $Z_\text t$ is  $M_\text t  = Z_\text t- 24$~\cite{34}. The number ``24'' in this formula comes from the number of completely occupied minority states which consist of one $s$, three $p$ and eight $d$ states and gives total 12 states \cite{34}. These Slater and Pauling rules join the electronic properties directly to the magnetic properties, and present a powerful tool to the study of HM Heusler compounds. The number of valence electrons $Z_\text t$ for Rh$_2$CrAl and Rh$_2$CrGe is 27 [$Z_\text t= (9 \times 2)+6+3=27$] and 28 [$Z_\text t= (9 \times 2) +6+4=28$], respectively, their total spin magnetic moments are $3.000$~$\muup \text B$ (integer value which confirms a half-metallic behavior) and $3.97579~\muup \text B$ (nearly integer which confirms a nearly half-metallic behavior), respectively.  The basis of the HM gap is discussed in herein below. The HM gaps frequently take place from three aspects~\cite{35}: 1)~charge transfer band gap \cite{35}  which is frequently seen in CrO$_2$ and double perovskites~\cite{35}, 2) covalent band gap which is present in the half-Heusler with $C1b$ structure, and  3) $d-d$ band gap, that is the origin of the HM band gap in the full-Heusler alloys with Cu2MnAl structure. In the latter case, there is a quite strong hybridization between $d$ orbitals of transition metals which makes the $d$-orbitals split into bonding $e_ \text g$ and $t_{2 \text g}$ orbitals below the Fermi level and anti-bonding $e_\text g^*$ and $t_{2 \text g}^*$   orbitals above the Fermi level. This hybridization is known as $d-d$ hybridization \cite{36}.

\begin{figure}[!t]
	\centering
	\subfloat[]{{\includegraphics[width=9cm]{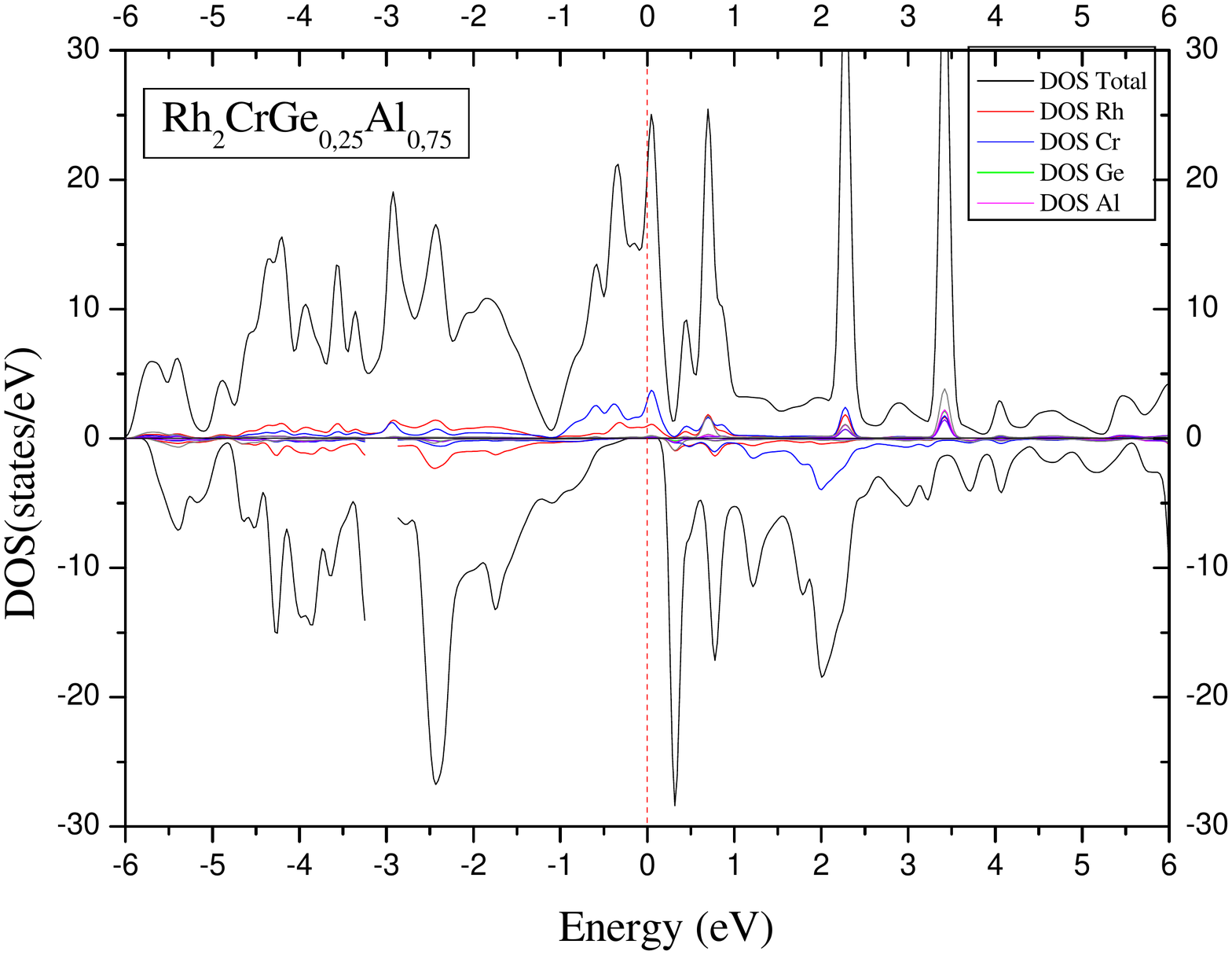} }}%
\vspace{-2mm}	
	\subfloat[]{{\includegraphics[width=9cm]{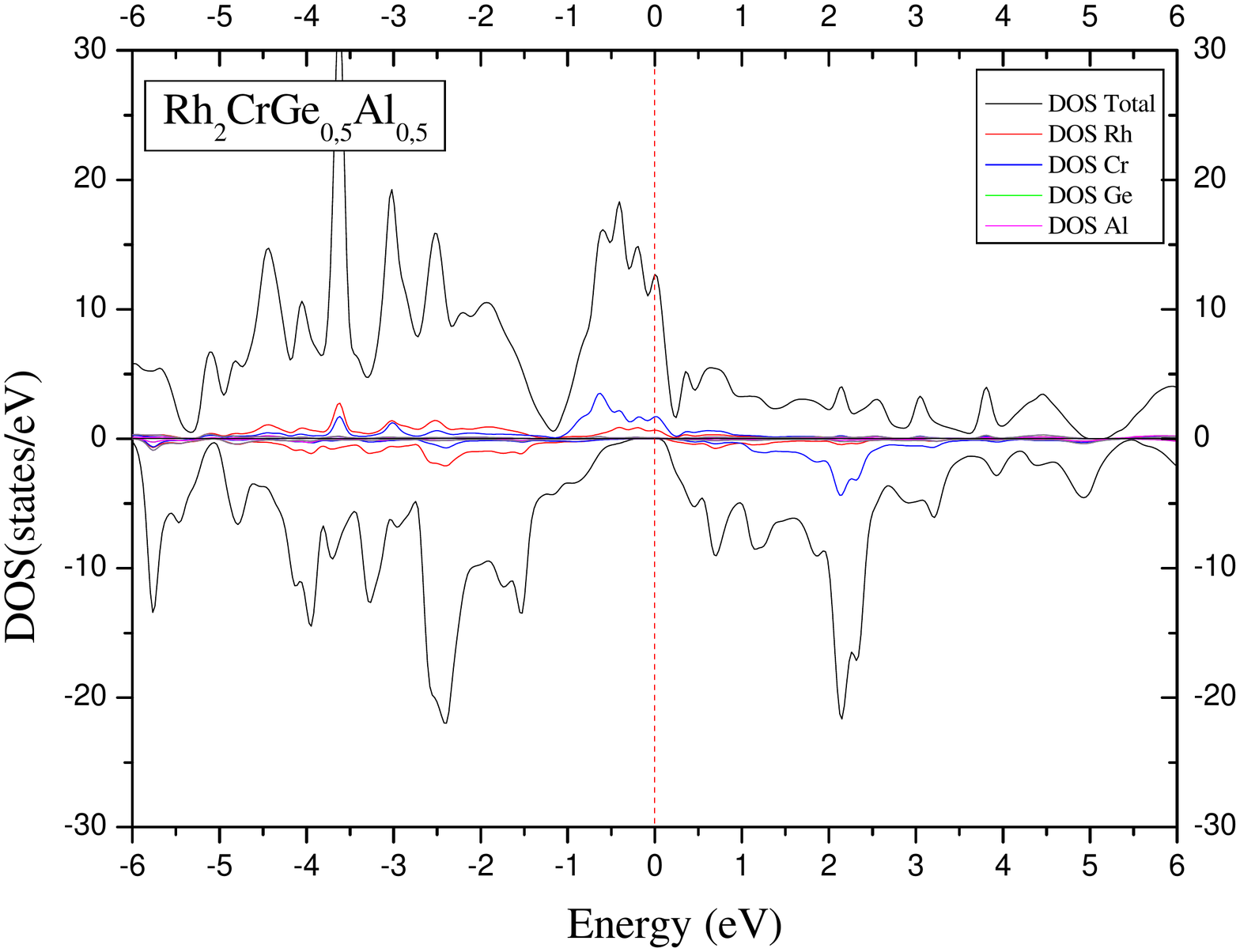} }}%
\vspace{-2mm}	
	\subfloat[]{{\includegraphics[width=9cm]{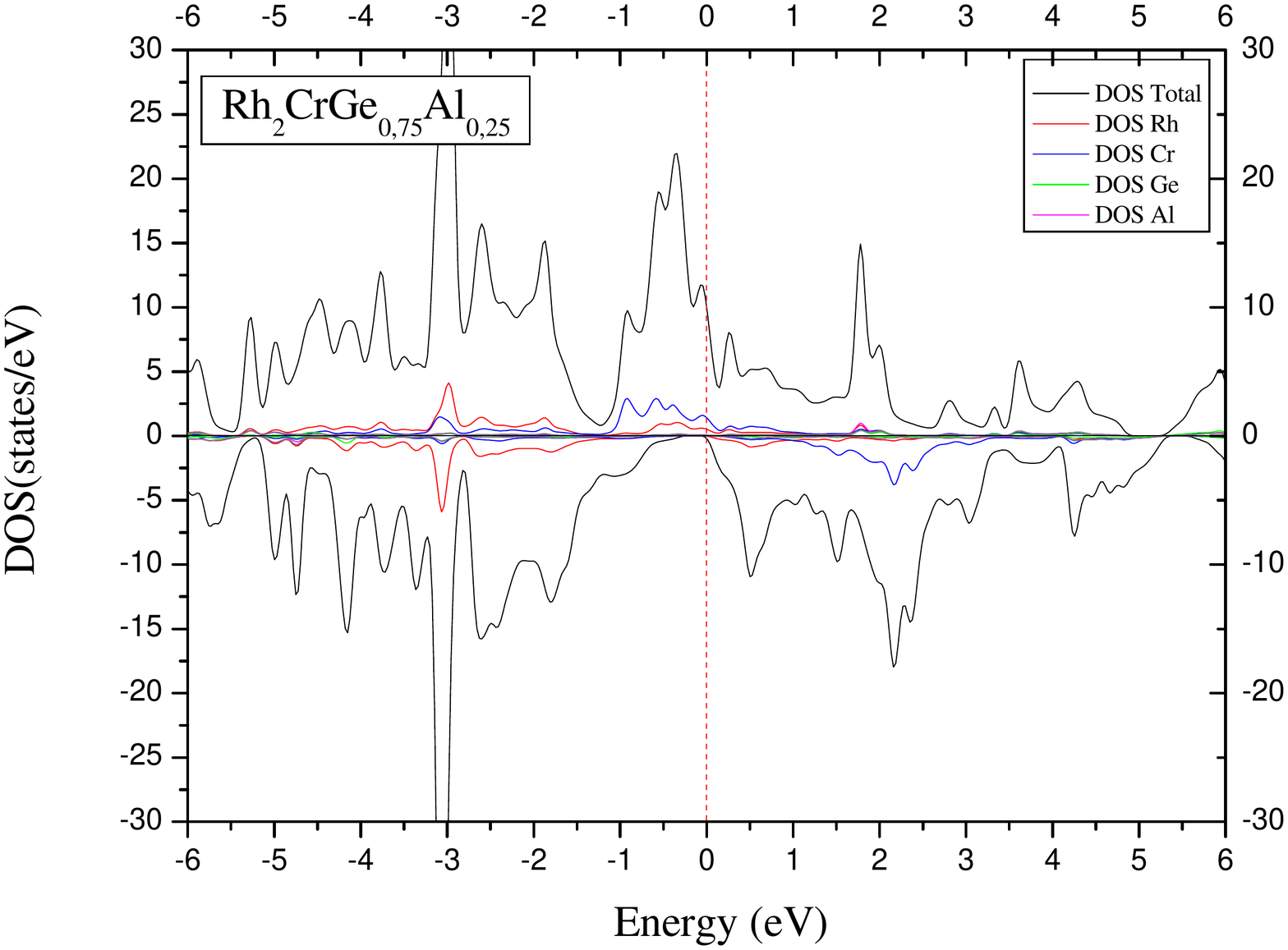} }}%
	\caption{(Colour online) Total and partial densities of states of Rh$_2$CrGe$_{1-x}$Al$_x$ ($x=0.25$, $0.5$ and $0.75$) compounds.}%
	\label{fig3}%
\end{figure}

The calculated total and partial densities of state of Rh$_2$CrGe$_{1-x}$Al$_x$ ($x =0.25$, $0.5$, $0.75$) compounds within the GGA approach are shown in figure~\ref{fig3}. One can observe the absence of the gap at Fermi level in the spin ($\uparrow$) and its presence in spin ($\downarrow$), which confirms the metallic behavior for spin ($\uparrow$) and the semi-conducting behavior for spin ($\downarrow$). As a result, the Rh$_2$CrGe$_{0.25}$Al$_{0.75}$, Rh$_2$CrGe{0.5}Al$_{0.5}$ and Rh$_2$CrGe$_{0.75}$Al$_{0.25}$ are full Heusler are half-metallic compounds. 
The DOS is characterized by a large domination of the Cr-$3d$ states, which leads to the large spin moments at their sites, which are around $12.99083~\muup \text B$, $14.00174~\muup \text B$ and $15.00682~\muup \text B$ for the Rh$_2$CrGe$_{0.25}$Al$_{0.75}$, Rh$_2$CrGe{0.5}Al$_{0.5}$ and Rh$_2$CrGe$_{0.75}$Al$_{0.25}$ compounds, respectively as shown in table~\ref{tab2}. It is seen that the total magnetic moment per unit cell decreases as a function of $x$ concentration ($4~\muup \text B$, $15.00682~\muup \text B /16$ atoms, $14.00174~\muup \text B/16$ atoms, $12.990~\muup \text B/16$ atoms, and $3~\muup \text B$, for $x   = 0, 0.25$, $0.50$, $0.75$, and  $1$, respectively),  for  Rh$_2$CrGe$_{1-x}$Al$_x$ ($x =0$, $0.25$, $0.5$, $0.75$ and $1$). The calculated spin magnetic moments for the Rh$_2$CrGe$_{0.25}$Al$_{0.75}$, Rh$_2$CrGe{0.5}Al$_{0.5}$ and Rh$_2$CrGe$_{0.75}$Al$_{0.25}$ alloys show that the total magnetic moment which includes the contribution from the interstitial region, comes mainly from the Cr ion with a small contribution of Rh sites. The Ge and Al atoms have a small anti-parallel spin moment to that of the Rh atom occupying the $X$ sites in the lattice. 

\begin{table}[!t]
\caption{The calculated total and partial magnetic moments (in$~\muup \text B$) for Rh$_2$CrGe, Rh$_2$CrAl and  Rh$_2$CrGe$_{1-x}$Al$_x$ ($x=0.25$, $0.5$ and $0.75$) compounds.}
\label{tab2}
\vspace{2ex}
\footnotesize
\begin{center}
\scriptsize
\renewcommand{\arraystretch}{1.3}
\begin{tabular}{|c|c||c|c|c|c|c|c|c||}
\hline
 & &$M$(Rh$_1$)& $M$(Rh$_2$)&$M$(Cr) &$M$(Ge)& $M$(Al)&$M_\text{int}$&$\muup$ Total \strut\\

\hline
 
  Rh$_2$CrGe&GGA-PBE &$0.35055$&$0.35055$ & $3.03681$&$-0.00748$& $-$ &0.24537&3.97579\strut\\

\hline
  
    Rh$_2$CrAl & GGA-PBE&0.19516&0.19516 &2.57815 &$-$0.0284& $-$0.06126&$-$ &3.000\strut\\

\hline
  \raisebox{-1.7ex}[0pt][0pt] {Rh$_2$CrGe$_{0.25}$Al$_{0.75}$}&\raisebox{-1.7ex}[0pt][0pt] {GGA-PBE}&\raisebox{-1.7ex}[0pt][0pt]{0.24261}&\raisebox{-1.7ex}[0pt][0pt]{0.24261}&\raisebox{-1.7ex}[0pt][0pt] {2.69551}&\raisebox{-1.7ex}[0pt][0pt]{$-$0.0059}&\raisebox{-1.7ex}[0pt][0pt]{$-$0.0256} &\raisebox{-1.7ex}[0pt][0pt]{0.34582} &12.99083\strut\\ 
  &&&& &&&& (with unit cell\strut\\ 
       &&&& &&&& of 16 atoms)\strut\\ 

\hline

  \raisebox{-1.7ex}[0pt][0pt]{Rh$_2$CrGe$_{0.5}$Al$_{0.5}$}& \raisebox{-1.7ex}[0pt][0pt]{GGA-PBE}&\raisebox{-1.7ex}[0pt][0pt]{0.29814}&\raisebox{-1.7ex}[0pt][0pt]{0.29814} &\raisebox{-1.7ex}[0pt][0pt]{2.81287}
 &\raisebox{-1.7ex}[0pt][0pt]{$-$0.0041}&\raisebox{-1.7ex}[0pt][0pt]{$-$0.0240}& \raisebox{-1.7ex}[0pt][0pt]{0.43532}&14.00174 \strut\\
  &&&& &&&& (with unit cell\strut\\ 
  &&&& &&&& of 16 atoms)\strut\\ 
\hline
\raisebox{-1.7ex}[0pt][0pt] {Rh$_2$CrGe$_{0.75}$Al$_{0.25}$}&\raisebox{-1.7ex}[0pt][0pt]{GGA-PBE}&\raisebox{-1.7ex}[0pt][0pt]{0.35669}&\raisebox{-1.7ex}[0pt][0pt]{0.35669} & \raisebox{-1.7ex}[0pt][0pt]{2.91783}&\raisebox{-1.7ex}[0pt][0pt]{$-$0.00014}&\raisebox{-1.7ex}[0pt][0pt]{$-$0.0237}
 & \raisebox{-1.7ex}[0pt][0pt]{0.51850}&15.00682 \strut\\
  &&&& &&&& (with unit cell )\strut\\ 
  &&&& &&&& of 16 atoms)\strut\\ 
    
\hline

\end{tabular}
\renewcommand{\arraystretch}{1}
\end{center}
\end{table}

\section{Conclusion}\label{sec4}
In this paper, we have presented theoretical results of structural, electronic and magnetic properties of Rh$_2$CrGe$_{1-x}$Al$_x$ Heusler alloys. Calculations were performed using the FP-LAPW method as implemented in WIEN2k code within GGA-PBE. The most important results are as follows: 

The alloys studied are more stable in Cu$_2$MnAl ($L21$) structure.

The ground state properties of the materials studied are determined in the two phases.

The calculated density of state of the ferromagnetic configuration for Rh$_2$CrGe and Rh$_2$CrAl show that the first one is nearly half-metallic and the second one is half-metallic.

For $x=0.25$, $0.50$ and $0.75$, the materials have a metallic behavior for spin-up and a semi-conducting behavior for spin-down, so these materials are half-metallic compounds which are most functional in spintronic. 

The magnetic moments were mostly contributed by the $3d$ orbital of Cr and $4d$ orbital of Rh ions.
In the absence of experimental and theoretical works, the present results for the alloys studied provide an estimate of these materials which can be useful for further studies.


%
%

\ukrainianpart

\title{Теоретичне дослідження фазової стійкості, електронних і магнітних властивостей 
	сплавів Хеслера  Rh$_2$CrGe$_{1-x}$Al$_x$ ($x = 0$, $0.25$, $0.50$, $0.75$ і $1$) методом  FP-LAPW }
\author{M. Каджауд\refaddr{label1}, M. Мохтарі\refaddr{label2,label3},
	Л. Джоуді\refaddr{label1,label3}, М. Мерабе\refaddr{label1,label3}, С. Беналья\refaddr{label1,label3}, Д. Рачед\refaddr{label1}, 
	Р. Белачел\refaddr{label1}, Ф. Замі\refaddr{label4}, Ф. Дахман\refaddr{label3}}
\addresses{
	\addr{label1} Лабораторія магнітних матеріалів, фізичний факультет, університет  Сіді-Бель-Аббес, Алжир
	\addr{label2} Університет природничих наук і технологій ім. Муххамеда Будіафа,  31000 Оран, Алжир
	\addr{label3} Інститут природничих наук і технологій, університетський центр  м. Тіссемсілт, 38000 Тіссемсілт, Алжир
	\addr{label4} Лабораторія конденсованої речовини і сталого розвитку,  фізичний факультет, університет  Сіді-Бель-Аббес, Алжир
}

\makeukrtitle

\begin{abstract}
	Проведено першопринципні обчислення в рамках теорії функціоналу густини  з використанням методу  лінеаризованих приєднаних
	плоских хвиль з повним потенціалом та коду WIEN2k з метою визначення структурної стійкості електронних і магнітних властивостей 
	Rh$_2$CrGe$_{1-x}$Al$_x$ ($x = 0$, $0.25$, $0.50$, $0.75$ і $1$). Результати показали, що для Rh$_2$CrAl і Rh$_2$CrGe, структура типу  Cu$_2$MnAl є енергетично більш стійкою, ніж структура типу  Hg$_2$CuTi при рівноважному об'ємі. Обчислені густини станів для Rh$_2$CrAl і Rh$_2$CrGe показали, відповідно,  напівметалеву та майже металеву поведінку. Rh$_2$CrGe$_{1-x}$Al$_x$ ($x = 0.25$, $0.50$, $0.75$) цих сплавів  показали напівметалеву поведінку. Ці сполуки можна вважати хорошими кандидатами для спінтронних застосувань.
	\keywords сплави Хеслера, структурні властивості, електричні властивості, магнітні властивості 
\end{abstract}


\end{document}